\documentclass[natbib]{emulateapj}
\submitted{MNRAS, published}
\usepackage{graphicx}
\usepackage{rotating}
\usepackage{amsmath}
\usepackage[usenames]{color}
\usepackage[colorlinks=true, pdfstartview=FitV, citecolor=blue, urlcolor= red]{hyperref}

\shortauthors{Oh et al.} \shorttitle{Robust profile decomposition for large extragalactic spectral-line surveys}

\begin{document}

\newcommand{\kms}{\ensuremath{\mathrm{km}\,\mathrm{s}^{-1}}}
\newcommand{\mjybeam}{\ensuremath{\mathrm{mJy}\,\mathrm{beam}^{-1}}}
\newcommand{\cm}{\ensuremath{\mathrm{cm}^{-2}}}
\newcommand{\kmsnospace}{\ensuremath{\mathrm{km}\,\mathrm{s}^{-1}}}
\newcommand{\barkms}{\ensuremath{\mathrm{km}\,\mathrm{s}^{-1}\,\mathrm{kpc}^{-1}}}
\newcommand{\kmsMpc}{\ensuremath{\mathrm{km}\,\mathrm{s}^{-1}\,\mathrm{Mpc}^{-1}}}
\newcommand{\etal}{et al.} \newcommand{\LCDM}{$\Lambda$CDM}
\newcommand{\MLmax}{\ensuremath{\Upsilon_{max}}}
\newcommand{\Lsun}{\ensuremath{\rm{{L}_{\odot}}}}
\newcommand{\Msun}{\ensuremath{\rm{{M}_{\odot}}}}
\newcommand{\mass}{\ensuremath{\rm{{\cal M}}}}
\newcommand{\magsq}{\ensuremath{\mathrm{mag}\,\mathrm{arcsec}^{-2}} }
\newcommand{\Lsundens}{\ensuremath{\rm L_{\odot}\,\mathrm{pc}^{-2}}}
\newcommand{\surfdens}{\ensuremath{\rm M_{\odot}\,\mathrm{pc}^{-2}}}
\newcommand{\cubedens}{\ensuremath{M_{\odot}\,\mathrm{pc}^{-3}}}

\newcommand{\hi}{H{\sc i} }


\title{Robust profile decomposition for large extragalactic spectral-line surveys}

\author{Se-Heon Oh\altaffilmark{1}, Lister Staveley-Smith\altaffilmark{2,3} \& Bi-Qing For\altaffilmark{2,3}}
\email{seheon.oh@sejong.re.kr}

\altaffiltext{1}{Department of Physics and Astronomy, Sejong University, 209 Neungdong-ro, Gwangjin-gu, Seoul 05006, Republic of Korea}
\altaffiltext{2}{International Centre for Radio Astronomy Research (ICRAR),
University of Western Australia, 35 Stirling Highway, Perth, WA 6009, Australia}
\altaffiltext{3}{The ARC Centre of Excellence for All Sky Astrophysics in 3 Dimensions (ASTRO 3D)}

\begin{abstract}
We present a novel algorithm which is based on a Bayesian Markov Chain Monte Carlo (MCMC) technique for performing robust profile analysis of a data cube from either single dish or interferometric radio telescopes.
It fits a set of models comprised of a number of Gaussian components given by the user to individual line-of-sight velocity profiles, then compares them and finds an optimal model based on the Bayesian Inference Criteria computed for each model.
The decomposed Gaussian components are then classified into bulk or non-circular motions as well as kinematically cold or warm components. The fitting based on the Bayesian MCMC technique is insensitive to initial estimates of the parameters, and suffers less from finding the global minimum in models given enough sampling points and a wide range of priors for the parameters.
It is found to provide reliable profile decomposition and classification of the decomposed components in a fully automated way, together with robust error estimation of the parameters as shown by performance tests using artificial data cubes. We apply the newly developed algorithm to the \hi data cubes of sample galaxies from the Local Volume H{\sc i} galaxy Survey (LVHIS). We also compare the kinematically cold and warm components, and bulk velocity fields with previous analyses made in a classical method. 
\end{abstract} 

\section{Introduction} \label{intro}

The interstellar medium (ISM) in galaxies is an important reservoir of gas for ongoing star formation. It is also an important sink for energy from outflows, supernovae (SNe), shock fronts, gravitational interactions, etc. These phenomena can result in gas displacement, heating and ionisation. An immediate consequence can be a broadening or skewing of line-of-sight velocity profiles  (\citealt{1997ApJ...490..710Y}; \citealt{1996ApJ...462..203Y}; \citealt{2003ApJ...592..111Y}). Even for isolated galaxies without any obvious tidal interactions, the ISM is vulnerable to internal hydrodynamical processes resulting from star formation. The ISM in dwarf galaxies even without significant star formation activities can be susceptible to turbulent processes (See e.g., \citealt{2004A&A...416..499V}, \citealt{2005ApJ...630..238D}).

Gas outflows which are driven by the deposition of energy (baryonic feedback) such as stellar winds and SNe will locally disturb the ambient ISM, and give rise to complex gaseous structures, complex kinematics and multiple phases (\citealt{2014MNRAS.445..581H}; \citealt{2016MNRAS.456..710F}). On the other hand, gravitational interaction between galaxies induces more systematic and symmetric large-scale morphological changes such as lopsideness and warping (\citealt{1995ApJ...447...82R}; \citealt{1981AJ.....86.1825B}; \citealt{1990ApJ...352...15B}). Correct quantitative analysis of the gaseous content of galaxies will allow for better separation of multiple environmental effects such as ram pressure stripping, tidal interaction, bars, and warps, and also allow for a better understanding of their past evolution.

On kpc or sub-kpc scales, gas outflows driven by star formation or SNe will contribute to turbulent random motion deviating from the underlying circular rotation of the host galaxy. The majority of holes or local cavities often found in \hi gas disk of galaxies could be from such local baryonic feedback processes (e.g., \citealt{2011AJ....141...23B}). The kinematical and morphological properties like expansion velocities, sizes and asymmetries can be used to infer the released energy. This in turn allows us to investigate the interplay between baryonic feedback and the ISM (\citealt{2010MNRAS.409.1088B}). Some gas clouds show substantial kinematic deviation from the rotation of the disk of host galaxies, up to several hundreds \kms\ above their projected velocity significantly deviating from the projected velocity at their positions, and are therefore classified as high-velocity clouds (HVCs; See e.g., \citealt{2018MNRAS.474..289W}). The origins of these HVCs could arise from accretion from the cosmic web, leftovers from evolution of the host galaxy, or galactic fountains (\citealt{1980ApJ...236..577B}; \citealt{2009ApJ...692..470M}; \citealt{2015MNRAS.447L..70F}). 

Gas clouds moving at anomalous velocities cause complex ISM structure and turbulent kinematics, which often distort the underlying H{\sc i} kinematics of the host galaxy. Previous studies have shown that modelling of such distorted motions is not straightforward when classical moment analysis is applied to velocity profiles which are asymmetric, non-Gaussian, or have multiple components. Parameters such as amplitude, central velocity and velocity dispersion become sensitive to estimation methodology. Such analyses can result in biased measurement for highly asymmetric non-Gaussian profiles. This causes uncertainties in deriving galaxy rotation curves, and for subsequent mass modelling.

To minimise the effect of asymmetric non-Gaussian velocity profiles in the derivation of representative properties, profile decomposition techniques using multiple components can be used, as implemented in several radio astronomical software packages such as GIPSY\footnote{Groningen Image Processing System \citep{1992ASPC...25..131V}}, AIPS\footnote{Astronomical Image Processing System}, and CASA\footnote{Common Astronomy Software Applications \citep{2007ASPC..376..127M}}.
However, even if a satisfactory model is found, further issues often remain: 1) fitting is invariably based on $\chi^{2}$ minimisation procedures, which are sensitive to initial parameter estimates; 2) fits that use many (typically more than three) components will often fail to converge; 3) even if the profile decomposition is successful, these packages do not permit automated classification of components or any easy way to identify spatially coherent velocity features.

\cite{Oh_2008, Oh_2011, 2015AJ....149..180O} performed multiple Gaussian decomposition of an \hi data cube of galaxies from THINGS\footnote{The \hi Nearby Galaxy Survey \citep{2008AJ....136.2563W}} and LITTLE THINGS\footnote{Local Irregulars That Trace Luminosity Extremes, The H{\sc i} Nearby Galaxy Survey \citep{2012AJ....144..134H}}, separating random non-circular motions from circularly rotating components using model reference velocity fields constructed from 2D tilted-ring analysis. However, their fitting method was based on a $\chi^{2}$ minimisation technique and still suffers from some of the above issues. Their method requires manual estimation of input parameters for each data set. Likewise, their method is limited in determining the optimal number of Gaussian components in a statistical robust manner. This makes it difficult to perform robust and automated profile analysis for a large number of \hi data cubes from large surveys like `Widefield ASKAP\footnote{Australian Square Kilometre Array Pathfinder \citep{2016PASA...33...42M}} L-Band Legacy All-sky Blind SurveY' (WALLABY; \citealt{2012PASA...29..359K}).

To circumvent the issues, we present a new profile decomposition technique based on Bayesian Markov Chain Monte Carlo (MCMC) statistics, and discuss its practical application to a number of sample galaxies from LVHIS\footnote{Local Volume \hi Survey \citep{2018MNRAS.478.1611K}}. We also present their bulk velocity fields which excludes the effects of non-circular motions as well as decomposed additional kinematic components. The newly developed algorithm will be useful for quantitative analyses of the structure and kinematics of the ISM in other galaxies with high resolution data, or for the automated analysis of data from future large \hi surveys.

The structure of this paper is as follows: Section~\ref{pd} describes the newly developed algorithm, and Section~\ref{pt} discusses a performance test using an artificial \hi data cube. Section~\ref{mp} presents a practical application to the LVHIS sample galaxies, and discusses the newly derived kinematics. Section~\ref{conc} summarises the main results.

\section{Profile decomposition}\label{pd}

\subsection{Modelling non-Gaussian velocity profiles} \label{pd1}

A non-Gaussian velocity profile at a line-of-sight velocity {\it x} can be modelled as a set of multiple Gaussian components as follows:
\begin{equation}\label{eq:1}
{G(x)} = \sum_{i=1}^{m}\frac{a_{i}}{\sqrt{2\pi}\sigma_{i}}\,\exp\left({\frac{-(x-\mu_{i})^2}{2\sigma_{i}^{2}}}\right) + \sum_{j=0}^{n}b_{j}\,x^{j} 
\end{equation}
\noindent where {\it m} is the number of Gaussian components fitted, $a_{i}$, $\sigma_{i}$, and $\mu_{i}$ are the area, standard deviation, and central velocity of the $i^{\rm th}$ Gaussian component. $b_{j}$ are constants of the $n^{\rm th}$-order polynomial for the baseline fit. The line-of-sight velocity profile of a single gas cloud element in the ISM of a galaxy is approximated by a Gaussian function. As given in Eq.~\ref{eq:1}, the line-of-sight velocity distribution of gas clouds moving at different velocities driven by internal or external hydrodynamical processes can then be described by multiple Gaussian components. 

As mentioned in Section~\ref{intro}, a fit made with multiple Gaussian components becomes sensitive to initial estimates of free parameters whose number, {\it M}, is proportional to the number ($m$) of Gaussian functions adopted (i.e., $3m$). This is particularly challenging for fitting algorithms based on a $\chi^{2}$ minimisation because they often get trapped in local minima of models. In addition, any spike-like noise properties in the profile worsen their ability to derive reliable estimates of the parameters. In this case, manual fitting in which initial estimates of parameters are provided by visual inspection is usually applied, however, supervision is usually required to achieve a global minimum. This is not ideal for profile analysis of large \hi surveys or even for a single observation. Moreover, it often leads to either under- or over-fitting due to the unknown number of Gaussian components in the prior.

In order to improve the situation, we perform a profile analysis in a Bayesian MCMC framework. Compared to a fitting algorithm based on a $\chi^{2}$ minimisation, it is more efficient at sampling high-dimensional parameter space like the case shown in Eq.~\ref{eq:1}. Despite a large CPU penalty, this algorithm is less likely to get trapped in local minima, and more able to find the global minima correctly, as long as there are enough sampling points and a sufficiently wide range of priors for each parameter. This significantly reduces the sensitivity of initial estimates of the $\chi^{2}$ fitting to the final fit results. As a bonus, the MCMC sampling technique provides robust error estimates of the parameters. Ultimately, this would allow us to perform robust and reliable profile decomposition analysis of data cubes from large spectral emission line surveys in an automated way.

We perform a Bayesian fitting of the model with multiple Gaussian components given in Eq.~\ref{eq:1} to a profile via MCMC sampling as follows,
\begin{equation} \label{eq:2}p(\Theta|y, g) = \frac{p(y|\Theta, g) \times
p(\Theta|g)}{p(y|g)}, \end{equation}
\noindent where $p(y|g)$ is the evidence, $p(\Theta|g)$ is the likelihood, $p(y|\Theta, g)$ is prior, and $p(\Theta|y, g)$ is the posterior distribution of the model from which we derive the best fits of the parameters (see \citealt{sivia2006data}). For the likelihood, we define a log-likelihood function for a Student-t distribution as follows,
\begin{align}\label{eq:3}
{\rm log}\,L &= \sum_{n=1}^{N}\,{\rm log}\biggl[\frac{\Gamma(\frac{\nu + 1}{2})}{\sqrt{\pi(\nu-2)}\Gamma(\frac{\nu}{2})}\biggr] \\ \nonumber
&- \frac{1}{2}\sum_{n=1}^{N}\,{\rm log}\sigma_{n}^{2}  \\ \nonumber
&- \frac{\nu+1}{2}\sum_{n=1}^{N}\,{\rm log}\biggl[1+\frac{\epsilon_{n}^{2}}{\sigma_{n}^{2}(\nu-2)}\biggr]
\end{align}
\noindent where $\epsilon_{n} = f_{n}^{\rm LOS} - G_{n}$ ($f_{n}^{\rm LOS}$: the observed flux at the $n^{\rm th}$ channel of a profile; $G_{n}$: a Gaussian model value at the $n^{\rm th}$ channel as given in Eq.~\ref{eq:1}), $N$ is the number of channels of a profile,
$\nu$ ($>$ 2) is the normality parameter, and $\Gamma$ is the gamma function given as
\begin{align}\label{eq:4}
\Gamma \left( x \right) = \int\limits_0^\infty {s^{x - 1} e^{ - s} ds} .
\end{align}
The overall scaling of the distribution is set by a free parameter, $\sigma_{n}$. As discussed in \cite{2018MNRAS.473.3256O}, the Student-t distribution is less sensitive to outlying data points by having heavier tails than the normal distribution (see also \citealt{2011PhRvD..84l2004R}). Following \cite{2018MNRAS.473.3256O}, we adopt $\nu=3$ which is the smallest value allowed in the Student-t distribution. 

\begin{figure*}
\includegraphics[angle=270,width=1.0\textwidth,bb=20 50 530 780,clip=]{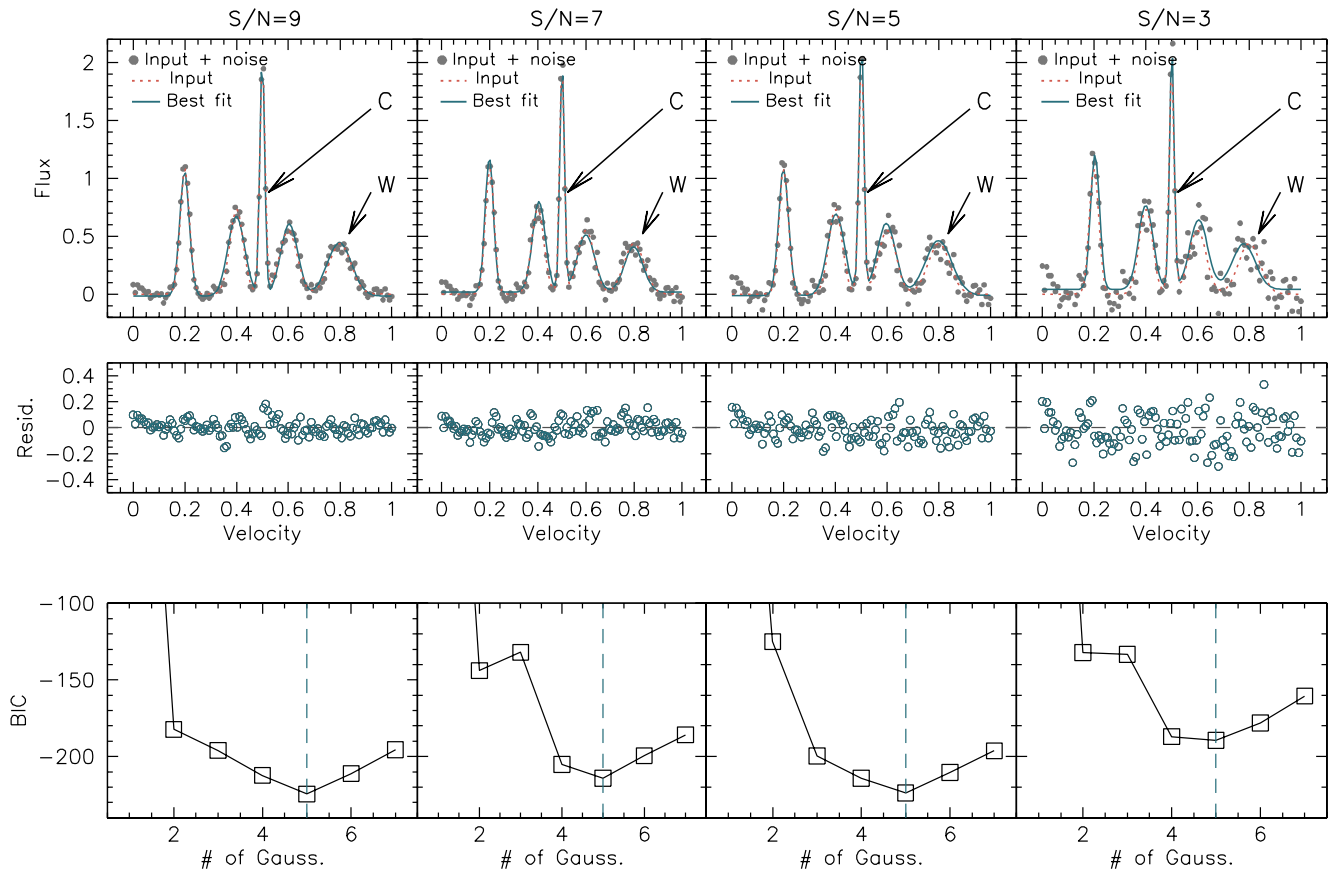}
\scriptsize
\caption{Example velocity profiles with different S/N values of 3, 5, 7 and 9. {\bf Upper panels}: Dotted lines indicate the input profiles and grey dots include added Gaussian noise. The solid lines show the best fits using the one with the least BIC value among the ones with different numbers of Gaussian components up to seven. The arrows indicate the kinematically cold (C) and warm (W) Gaussian components classified. {\bf Middle panels}: Residuals between the input noise-added profiles and the best fits in the upper panels. {\bf Bottom panels}: The computed BIC values in Eq.~\ref{eq:7} from the Bayesian fits with different number of Gaussian components from 1 to 7. The dashed lines indicate the number of Gaussian components with which the Bayesian fits have the least BIC values.
\label{Fig1}}
\end{figure*}

For the calculation of the Bayesian evidence in Eq.~\ref{eq:2}, which is the most time consuming step in a Bayesian analysis, we use the {\sc multinest} library \citep{2008MNRAS.384..449F,2009MNRAS.398.1601F} which implements a nested sampling algorithm. {\sc multinest} is a Bayesian inference tool which has been applied to several astrophysical inference problems with high multi-modality (e.g., \citealt{2009CQGra..26u5003F}; \citealt{2011MNRAS.415.3462F}). It is particularly useful for parameter estimation and model selection in multi-modal posterior distributions as in our case. Similar implementation has been applied to absorption line data from the ASKAP `First Large Absorption Survey in H{\sc i}' (FLASH) \citep{2012PASA...29..221A}. Most recently, it has been successfully implemented in a 2D galaxy kinematics analysis tool, {\sc 2dbat} (\citealt{2018MNRAS.473.3256O}), and found to be robust and efficient in the kinematic parameterisation of 26 resolved galaxies from LVHIS. 

For the prior distributions of the parameters in Eq.~\ref{eq:2}, we use conservative uniform priors which are wide enough to cover their possible ranges and to avoid any biases caused by priors. We normalise the amplitudes of profiles in a data cube to make their priors unitless as follows,
\begin{align}\label{eq:5}
f_{\rm norm}(x_{\rm norm}) = \frac{f(x_{\rm norm}) - f_{\rm min}}{f_{\rm max} - f_{\rm min}}
\end{align}
\noindent where $f_{\rm min}$ and $f_{\rm max}$ are the lowest and highest flux values of a profile, and $x_{\rm norm}$ is a normalised velocity in the same way,
\begin{align}\label{eq:6}
x_{\rm norm} = \frac{x - x_{\rm min}}{x_{\rm max} - x_{\rm min}}
\end{align}
\noindent where velocity {\it x} ranges from $x_{\rm min}$ to $x_{\rm max}$.

For an objective and quantitative model selection among the other competing models having different $N$ in Eq.~\ref{eq:1}, we use the Bayesian Information Criteria (BIC) value which is computed as follows:
\begin{align}\label{eq:7}
{\rm BIC} = \log(N)\,p - 2\,\log(\hat{L})
\end{align}
\noindent where $N$ is the number of data points, {\it p} is the number of free parameters to fit, and $\hat{L}$ is the maximised likelihood function of the model. 
This evaluates the statistical benefit of adding additional free parameters to the fit. As the number of Gaussian components $m$ adopted in Eq.~\ref{eq:1} increases, the maximised value of the likelihood function tends to increase, favouring a model with the larger Gaussian components $m$. However, the number of free parameters $p$, in the first term of Eq.~\ref{eq:7} acts as a penalty against over-fitting, balancing between the high-likelihood and the model complexity. In this way, for a finite set of models with different number of Gaussian components $m$ in Eq.~\ref{eq:1}, we can choose a model with the lowest BIC value as the best model for a given profile.

\subsection{Separation and classification of the decomposed multiple Gaussian components} \label{pd2}

The individual Gaussian components in a model selected according to BIC values are then classified into sub-components in accordance with their kinematic properties as follows: 1) the bulk motion moving at a velocity nearest the {\it assumed} underlying kinematics of the host galaxy; 2) non-circular motions deviating from the bulk motion; 3) kinematically cold or warm components with either smaller or larger velocity dispersion; 4) high-velocity components moving at a velocity larger than the {\it allowed} galactic velocity in the same direction.

For the {\it assumed} underlying kinematics of the host galaxy, a model velocity field from a 2D or 3D kinematic analysis is used. As an example, ring parameters from 2D tilted-ring analysis made to a velocity field can be used for constructing such a reference velocity field. To minimise dependence on the reference velocity field, which might result in biased bulk motions as well as other sub-components, we fit a simple disk model with constant position angle ($\phi$) and inclination ($i$) to smooth out any localised kinematic deviations and better extract the underlying global kinematics. In addition, as discussed in \cite{Oh_2008}, for constructing a reference velocity field, we start with a single Gaussian fit, or moment 1, to which a 2D or 3D disk model is fitted. We then iterate the procedure by replacing the reference velocity field with a new one constructed from the kinematic analysis done using the bulk velocity field extracted in the previous step. We stop the process once the average of velocity residuals between the successive bulk velocity fields extracted is less than a certain value, such as the channel resolution. Lastly, we separate non-circular motions from the circular rotation of the galaxy based on the extracted bulk velocity field.

\begin{figure*}
\includegraphics[angle=0,width=1.0\textwidth,bb=-20 340 610 750,clip=]{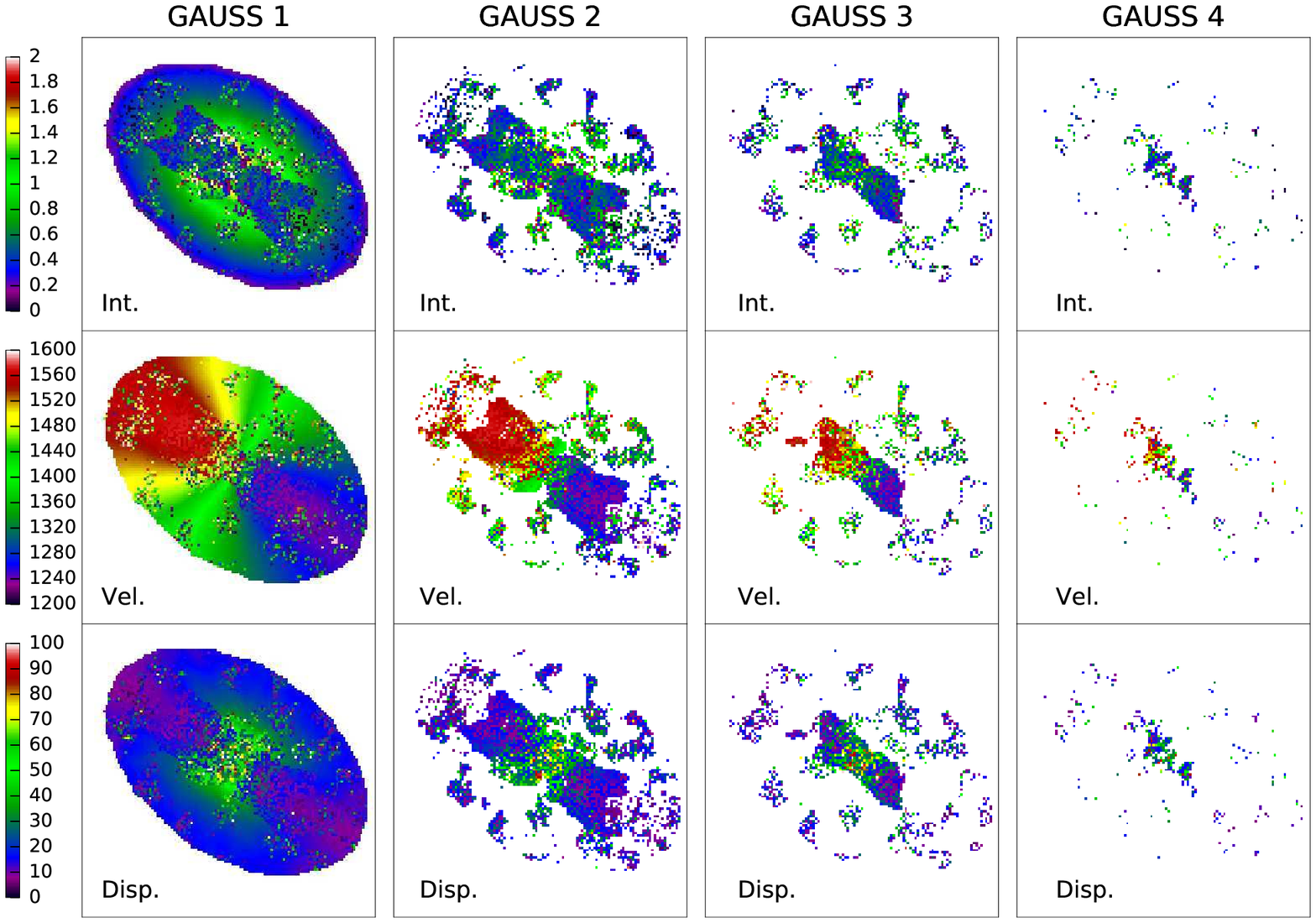}
\scriptsize
\caption{Example 2D maps of the four Gaussian components extracted from an artificial data cube described in Section~\ref{pt1}: 1) Int. (integrated intensity in units of $\rm Jy\,\rm beam^{-1}\,\rm km\,\rm s^{-1}$), 2) Vel. (central velocity in units of \ensuremath{\mathrm{km}\,\mathrm{s}^{-1}}), and 3) Disp. (velocity dispersion in units of \ensuremath{\mathrm{km}\,\mathrm{s}^{-1}}). Note that these are not sorted in intensity, central velocity or velocity dispersion.
\label{Fig2}}
\end{figure*}

\subsection{Software} \label{pd3}

We developed a standalone computer code written in C that implements the decomposition of velocity profiles in a data cube and subsequent classification as described in Sections~\ref{pd1} and \ref{pd2}. This software employs {\sc multinest}, a Bayesian inference tool, for calculating the posterior distribution and the evidence for a given likelihood function. Additional libraries like CFITSIO \citep{1999ASPC..172..487P}, GNU Scientific Library (GSL) and some routines from Numerical Recipes \citep{numerical_recipe} are used in the course of the analysis. 

To improve the processing time in calculating the evidence using MCMC techniques, the code is parallelised using the Message-Passing Interface (MPI) standard. In the current version, the input data cube is equally segmented into a number of cubelets to which each processor of a multi-core machine is assigned. In this way, the execution time can be reduced by using multiple processors.
The code is available for download\footnote{http://www.github.com/seheonoh/baygaud}, alongside instructions for its installation.

\subsection{Examples} \label{pd4}

In this section, we demonstrate the robustness and reliability of the method using example velocity profiles in a variety of situations with different noise properties and complexity. For complexity, we simulate a velocity profile consisting of up to four additional Gaussian components defined with their own amplitudes, central velocities and dispersions. We then add different levels of Gaussian noise to the profiles. The resulting signal-to-noise ratio (S/N) values of their lowest Gaussian components are 3, 5, 7, and 9 as presented in the upper panels of Fig.~\ref{Fig1}.

We fit a set of models given in Eq.~\ref{eq:1} to the profiles with $m$=1 to 7 to assess whether the fit results are reliable and robust. The Gaussian components decomposed using the code are overplotted as dotted lines in the upper panels of Fig.~\ref{Fig1}.
As shown in the bottom panels of Fig.~\ref{Fig1}, for all the example profiles, the number of input Gaussian components, namely five, is correctly recovered. In addition, the derived parameters of the individual Gaussian components are in good agreement with the input ones, within uncertainties, even for S/N $\sim$3. This is also apparent in the residual profiles shown in the middle panels of Fig.~\ref{Fig1}. Additionally, the kinematically cold and warm Gaussian components, classified according to their velocity dispersions are correctly indicated by arrows in the upper panels of Fig.~\ref{Fig1}. 

\begin{figure*}
\includegraphics[angle=0,width=1.0\textwidth,bb=0 470 600 780,clip=]{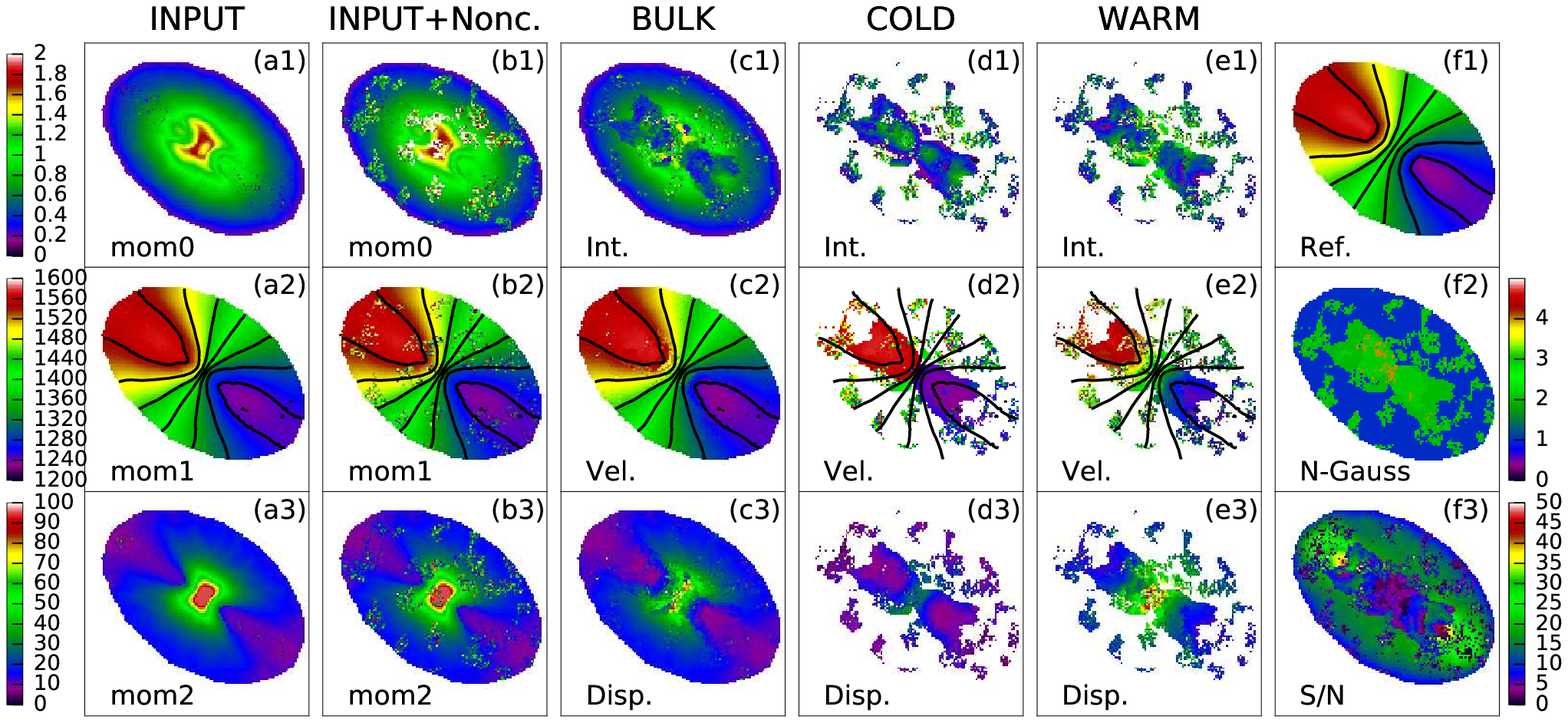}
\scriptsize
\caption{Example 2D maps of an artificial data cube described in Section~\ref{pt1}. The Gaussian components classified as bulk motions, kinematically cold or warm components as well as the input moments with and without non-circular motions. The moment 0 and integrated intensity (Int.) maps are given in units of $\rm Jy\,\rm beam^{-1}\,\rm km\,\rm s^{-1}$. The moment 1, 2, central velocity (Vel), and velocity dispersion (Disp.) maps are given in units of \ensuremath{\mathrm{km}\,\mathrm{s}^{-1}}. The maps denoted as Ref., N-Gauss, and S/N in the rightmost panels show 1) the reference velocity field used for extracting the bulk motions, 2) the optimal number of Gaussian components derived for each spaxel whose S/N values are larger than 3, and 3) the highest flux values of the best fits over the rms, respectively. See Section~\ref{pt2} for more details.
\label{Fig3}}
\end{figure*}

\begin{figure*}
\includegraphics[angle=0,width=1.0\textwidth,bb=10 190 600 620,clip=]{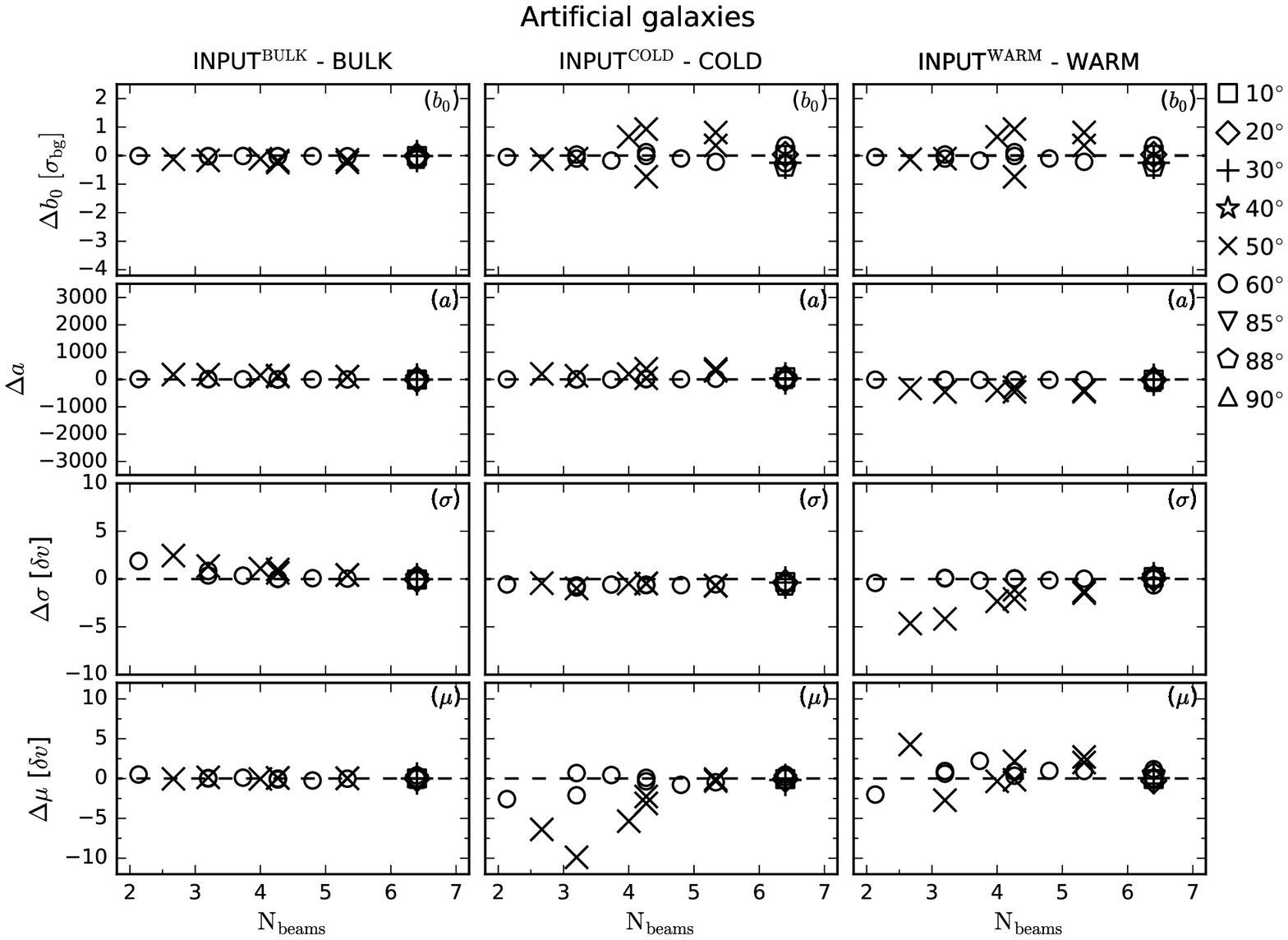}
\scriptsize
\caption{Mean residuals of the profile fit results ($b_{0}$: background, $a$: amplitude, $\sigma$: dispersion and $\mu$: central velocity) of the artificial cubes between the input and derived ones against the number of resolved elements $\rm N_{\rm beams}$ across the semi-major axis ($\rm 2 < N_{beams} < 7$). These are given in unit of the rms of $b_{0}$ ($\sigma_{\rm bg}$) and the channel resolution ($\delta v$). Different symbols indicate the range (10$^{\circ}$ \-- 90$^{\circ}$) of inclinations used for modelling the cubes in Section~\ref{pt1} \-- {\bf 1)} $\rm INPUT^{BULK}-BULK$: the difference between the input circular motions and the derived bulk motions; {\bf 2)} $\rm INPUT^{COLD}-COLD$: the difference between the input non-circular motions with narrower velocity dispersions and the motions classified as kinematically cold components; {\bf 3)} $\rm INPUT^{WARM}-WARM$: the difference between the input non-circular motions with wider velocity dispersions and the motions classified as kinematically warm components. See Section~\ref{pt2} for more details.
\label{Fig4}}
\end{figure*}

\begin{figure*}
\includegraphics[angle=0,width=1.0\textwidth,bb=10 190 600 620,clip=]{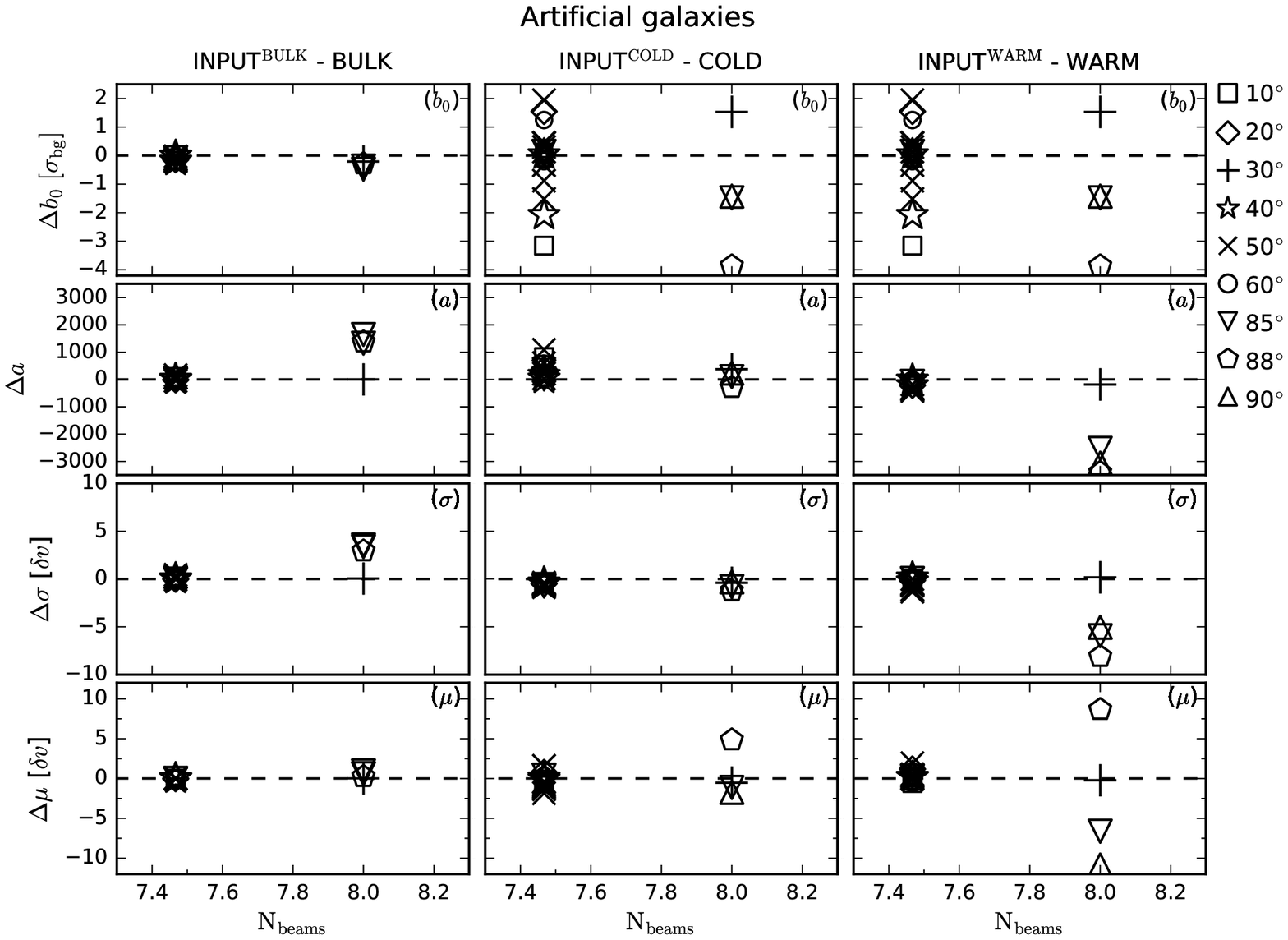}
\scriptsize
\caption{Mean residuals of the profile fit results ($b_{0}$: background, $a$: amplitude, $\sigma$: dispersion and $\mu$: central velocity) of the artificial cubes between the input and derived ones against the number of resolved elements $\rm N_{\rm beams}$ across the semi-major axis ($\rm 7 < N_{beams} < 9$). These are given in unit of the rms of $b_{0}$ ($\sigma_{\rm bg}$) and the channel resolution ($\delta v$). Different symbols indicate the range (10$^{\circ}$ \-- 90$^{\circ}$) of inclinations used for modelling the cubes in Section~\ref{pt1} \-- {\bf 1)} $\rm INPUT^{BULK}-BULK$: the difference between the input circular motions and the derived bulk motions; {\bf 2)} $\rm INPUT^{COLD}-COLD$: the difference between the input non-circular motions with narrower velocity dispersions and the motions classified as kinematically cold components; {\bf 3)} $\rm INPUT^{WARM}-WARM$: the difference between the input non-circular motions with wider velocity dispersions and the motions classified as kinematically warm components. See Section~\ref{pt2} for more details.
\label{Fig5}}
\end{figure*}

\section{Performance test} \label{pt}

In this section, we apply the technique to a set of model \hi data cubes of artificial galaxies whose underlying kinematics resembles typical rotation curves of intermediate-mass and massive spiral galaxies as well as dwarf galaxies. This is to verify the robustness and reliability of the algorithm described in Section~\ref{pd}, and the performance of the code.

\subsection{Artificial cubes} \label{pt1}

As base models, we adopt the 52 artificial data cubes which were used to test {\sc fat}\footnote{Fully Automated TiRiFiC \citep{2015MNRAS.452.3139K}} described in \cite{2015MNRAS.452.3139K}.
These were also used to test the performance of {\sc 2dbat} in \cite{2018MNRAS.473.3256O}. 
The cubes were constructed using three representative rotation curves, 1) a solid body-like rotation curve of dwarf galaxies, 2) an increasing rotation curve of intermediate-mass galaxies which is steeper than the solid body-like one but with no sign of becoming flat, and 3) a flat rotation curve of massive spiral galaxies. The H{\sc i} line flux is then distributed in the cubes based on their surface brightness profiles which are described by an exponential profile with a scale length varying with the size of the model galaxies. The cubes are spaced by a channel resolution of 4 \kms\ and smoothed by a Gaussian beam with FWHM of 30\arcsec\ after adding white noise. The spatial and spectral resolutions of the cubes are comparable to those of ASKAP observations at 21cm. The 52 base cubes cover a broad range of galaxy and observational properties such as geometry (inclination, position angle, and scale height), kinematics (warps, velocity dispersion, angular momentum vector, and rotation curves) and S/N values. The variations of some key parameters of the model cubes are listed in Table~\ref{table:1}. We refer to \cite{2015MNRAS.452.3139K} for a complete description of the model cubes.

\begin{table}
\caption{The values of some key parameters of the model cubes. See \cite{2015MNRAS.452.3139K} for more details.}
\label{table:1}
\scriptsize
   \centering
   \begin{tabular}{@{} lc @{}}
   \hline
\hline
Parameter    & values \\
\hline
The number of beams across the major axis & 4, 6, 7, 8, 9, 10, 11, 12, 16 \\
Inclination ($^{\circ}$) & 10, 20, 30, 40, 50, 60, 85, 88, 90 \\
Position angle ($^{\circ}$) & 45, 55 \\
Channel width (\kms) & 4 \\
Signal to noise ratio & 2, 4, 8, 16 \\
\hline
\end{tabular}
\end{table}

We randomly add additional Gaussian components to each line profile of the base data cubes to mimic random non-circular motions caused by SNe explosions and stellar winds from young stars. To explore an extreme case, we give the profiles in the data cube a large fraction of non-Gaussianity larger than 70\%. For simplicity, we add two different Gaussian components whose velocity dispersion is narrower (kinematically {\it cold}) or wider (kinematically {\it warm}) than the parent profile. This results in asymmetric non-Gaussian velocity profiles. We note that the usage of Gaussian profile could be limited in modelling any non-Gaussian velocity profiles of galaxies with high inclinations. The overlapped gas clouds being located at the same line-of-sight but moving at different rotation velocities could result in non-Gaussian velocity profiles, which becomes more prominent in edge-on-like galaxies. However, any inhomogeniety of gas distribution in the disk of these edge-on-like galaxies makes it difficult to predict a base model of their line-of-sight velocity profiles. Despite this limitation, we adopt the Gaussian profile as a base model for the line-of-sight velocity profiles of the disk. This will be more discussed in Section~\ref{pt2}.

We assign the Gaussian profiles of the random motions with their own amplitudes ($a$), velocity dispersions ($\sigma$), and velocity centres ($\mu$). We generate these parameters ($a$, $\sigma$, $\mu$) using Monte-Carlo techniques. 
For this, we randomly shift the central velocity of each parent line profile, and vary its amplitude, ranging from $\mu^{parent} -5\sigma^{parent}$ to $\mu^{parent} + 5\sigma^{parent}$, and $0.2 \times a^{parent}$ to $2.0 \times a^{parent}$ in the Monte Carlo simulations, respectively. For the velocity dispersion, we use two different ranges: 1) $0.2 \times \sigma^{parent}$ $\sim$ $0.7 \times \sigma^{parent}$, and 2) $1.5 \times \sigma^{parent}$ $\sim$ $2.0 \times \sigma^{parent}$ for the kinematically cold and warm components. Although this is not physically motivated, the resulting non-Gaussian velocity profiles are suitable enough to test the performance of our profile decomposition algorithm.

\subsection{Multiple Gaussian components} \label{pt2}

We run the code on the artificial \hi data cubes to assess whether it correctly decomposes the profiles and whether the input parameters are recovered. As introduced in Section~\ref{pd1}, we apply {\sc 2dbat} to moment 1 of the cube to derive an initial reference velocity field which is then updated in the next round of profile decomposition. 
For a given input velocity field, {\sc 2dbat} performs Bayesian fits of 2D tilted-ring models and automatically extracts ring parameters given a degree of regularisation for $\phi$ and $i$. We refer to \cite{2018MNRAS.473.3256O} for its full description and performance test.

When running {\sc 2dbat}, we assume constant $\phi$ and $i$ to best model the flat disc kinematics of the artificial galaxy. After profile decomposition according to BIC statistics, the initial bulk velocity field is extracted in accordance with the initial reference velocity field. We then re-run {\sc 2dbat} on the bulk velocity field to derive a new reference velocity field. This step is iterated until the mean residuals between the successive bulk velocity fields are less than the channel resolution of 4.0 \ensuremath{\mathrm{km}\,\mathrm{s}^{-1}}.

We performed the profile decomposition of the 52 artificial galaxies using the code, and classified the decomposed Gaussian components with respect to either their velocity dispersions and amplitudes or their reference velocity fields derived using {\sc 2dbat}. When performing the analysis, we let the code fit a given velocity profile with up to four Gaussian components simultaneously. This is more than enough to model the extreme cases of non-Gaussianity in the line profiles.  Accordingly, as described in Section~\ref{pd1}, the code makes four Bayesian fits of Eq.~\ref{eq:1} to each profile with $m$=1, to 4, and chooses the most appropriate model for the profile based on the computed BIC statistics. For instance, in Fig.~\ref{Fig2}, we show all the four decomposed Gaussian components of a model data cube where two additional Gaussian components are included. In addition, the extracted maps of the bulk motions, non-circular motions, kinematically cold and warm gas components are presented in Fig.~\ref{Fig3}.

As shown in the panel (f2) of Fig~\ref{Fig3}, the number of decomposed Gaussian components as per the BIC statistics is three in the regions where the lowest S/N value of the decomposed Gaussian components is larger than approximately 3. We remind the reader that the input number of Gaussian components used when constructing the \hi data cube in Section~\ref{pt1} was three. Moreover, the extracted bulk velocity field is similar to the one shown in the panel (a2) which does not include non-circular motions. In addition, the kinematically cold and warm velocity components are correctly separated given their velocity dispersions if the number of decomposed Gaussian components is at least larger than 2. 

From this, the flat disk model with constant $\phi$ and $i$ is likely to be a good approximation for the underlying kinematics of this artificial galaxy. However, this will not hold for severely disturbed galaxies, for instance, the ones experiencing strong tidal interactions. For these, a fine-tuned 2D tilted-ring analysis with a higher-order regularisation of $\phi$ and $i$ would be required in order to make a more representative reference velocity field. 

In Figs.~\ref{Fig4} and \ref{Fig5}, we compare the derived parameters ($bg$, $a$, $\sigma$ and $\mu$) of optimally extracted Gaussian components of all the model cubes with those that were used to build the cubes. For each model galaxy, we calculate the mean difference ($\Delta b_{0}$, $\Delta a$, $\Delta \sigma$ and $\Delta \mu$) between the derived and input Gaussian parameters of all the line profiles whose S/N value is larger than 3. The calculated mean difference of the model cubes is presented against the numbers of resolved elements across their semi-major axes, $\rm N^{\rm semi-mx}$. 
Different symbols indicate the input inclination values of the model galaxies from $10^{\circ}$ to $90^{\circ}$ as denoted in the plot. Different colours represents the beam sizes grouped into the bins from 2 to 8 beams. In addition, the comparison is made for the three Gaussian components classified as `bulk', `cold' and `warm'.

For the comparison between the bulk and input Gaussian components, the parameters recovered by the code are in general good agreement with the input except for the galaxies which are resolved by less than four beams across the semi-major axis and whose inclinations are larger than $70^{\circ}$ in our test. As expected, these deviations can be mainly attributed to the low-resolution beam smearing and the projection effects, respectively.
A similar trend is also found for the cold and warm Gaussian components although the offsets are slightly larger. The offsets are most prominent in the edge-on-like galaxies with $i > 80^{\circ}$ which are affected by projection, regardless of $\rm N^{\rm semi-mx}$. However, even in galaxies with intermediate inclinations, the beam smearing caused by poor sampling can also lead to large offsets in the derived Gaussian parameters. 
This demonstrates a fundamental limitation of profile decomposition for either edge-on or poorly-sampled galaxies. 

Meanwhile, for the well-resolved ($\rm N^{\rm semi-mx} > 4$) artificial galaxies with inclinations less than $60^{\circ}$, the Gaussian parameters recovered by the code in a fully automated manner are consistent with the input, and are correctly separated and classified in accordance with their velocity dispersions and reference velocity fields.

In summary, for galaxies affected by non-circular motions but still dominated by circular rotation, the code enables us to decompose different kinematic components and classify them correctly, as long as 1) their inclinations are less than $60^{\circ}$, 2) the number of resolved elements across their semi-major axes, $\rm N^{\rm semi-mx}$ is larger than four, and 3) the S/N value of profiles in their \hi data cubes is high enough to perform reliable profile decomposition.

\begin{table*}
\caption{Properties of the LVHIS sample galaxies}
\label{table:2}
\scriptsize
   \centering
   \begin{tabular}{@{} llcrrrrrrrr @{}}
   \hline
\hline
      Galaxy    & $\alpha$ (J2000) & $\delta$ (J2000) & H{\sc i} flux & $\rm log M_{HI}$ & $w_{50}$ & $v_{\rm SYS}$ & $\langle \rm P.A. \rangle$ & $\langle \rm i \rangle$ & $D$ & $\Theta_{\rm MX}$ \\
           &  (hh:mm:ss) & (dd:mm:ss) & ($\rm Jy\,\kms$) & ($\rm M_{\odot}$) & (\kms) & (\kms) &  ($^\circ$) & ($^\circ$) & (Mpc) & \\
        (1) & (2) & (3) & (4) & (5) & (6) & (7) & (8) & (9) & (10) & (11) \\
\hline
HIPASS J0256-54 &02 56 55 &-54 34 58 &	139.2	&	9.01	&	122	&	 574$\pm$2	  &  220 & 67 &6.8 & 13.30 \\
HIPASS J0320-52 &03 20 05 &-52 11 34 &	14.6 	&	7.97	&	80	&	 568$\pm$5	  &	 43  & 47 &5.3 & 3.69 \\
HIPASS J0705-58 &07 05 18 &-58 31 19 &	34.8 	&	8.29	&	68	&	 564$\pm$2	  &  277 & 28 &4.9 & 5.63 \\
HIPASS J1057-48 &10 57 32 &-48 11 02 &	104.4	&	8.83	&	67	&	 598$\pm$2	  &  120 & 32 &5.3 & 11.86 \\
HIPASS J1428-46 &14 28 06 &-46 18 32 &	17.3 	&	7.72	&	48	&	 390$\pm$2	  &  116 & 40 &3.4 & 4.88 \\
HIPASS J1620-60 &16 20 56 &-60 29 18 &	37.4 	&	8.56	&	139	&	 605$\pm$3	  &  33  & 51 &5.9 & 5.35 \\
\hline
\end{tabular}
\begin{minipage}{170mm}
\scriptsize{
{\bf (1):} HIPASS names;
{\bf (2)(3):} Kinematic centre positions derived from tilted-ring analyses in \cite{2018MNRAS.473.3256O}.
{\bf (4):} H{\sc i} flux density derived in \cite{2018MNRAS.478.1611K}.
{\bf (5):} Total H{\sc i} mass derived in \cite{2018MNRAS.478.1611K}.
{\bf (6):} H{\sc i} velocity widths determined at 50\% of the H{\sc i} peak flux in \cite{2018MNRAS.478.1611K}. 
{\bf (7):} Systemic velocity derived from tilted-ring analyses in \cite{2018MNRAS.473.3256O}.
{\bf (8):} Mean position angle derived from the tilted-ring analysis in \cite{2018MNRAS.473.3256O}.
{\bf (9):} Mean inclination derived from the tilted-ring analysis in \cite{2018MNRAS.473.3256O}.
{\bf (10):} Distance as given in \cite{2018MNRAS.478.1611K}.
{\bf (11):} Number of beams across the morphological major axis in \cite{2018MNRAS.473.3256O}}.
\end{minipage}
\end{table*}

\section{Application to the LVHIS sample galaxies} \label{mp}

We make a practical application of the method described in Section~\ref{pd} to a sample of galaxies taken from LVHIS \citep{2018MNRAS.478.1611K}. LVHIS is a southern sky H{\sc i} survey for 82 nearby ($<$ 10 Mpc) galaxies with the Australia Telescope Compact Array (ATCA) which provides a comprehensive H{\sc i} galaxy atlas in tandem with deep Anglo-Australian Telescope H-band images as well as {\it BVRI-}, H$\alpha$-, {\it GALEX uv-} and {\it Spitzer} mid-infrared images \citep{2018MNRAS.478.1611K}. One of its primary goals is to derive H{\sc i} morphologies and kinematics of the sample galaxies at high angular resolution and sensitivity, and relate them with star formation activities and environments. Thus, analysis of the H{\sc i} line profiles in an appropriate and quantitative manner is an essential prerequisite for addressing the structure and kinematics of the ISM, and the link to other hydrodynamical processes in the galaxies. In this paper, we focus on the performance of the code implementing the proposed profile decomposition method described in section~\ref{pd1} to see whether or not it is applicable to data cubes from H{\sc i} galaxy surveys. A companion paper will discuss the detailed kinematic analysis of the LVHIS galaxies (Oh et al. in prep).

Of the sample galaxies, we select 6 representative galaxies whose inclination and $\rm N^{\rm semi-mx}$ values are comparable to those of the artificial galaxies discussed in Section~\ref{pt1}: 1) 3 $<$ $\rm N^{\rm semi-mx}$; 2) $20^{\circ} < i < 70^{\circ}$. The basic observational properties of the sample galaxies are presented in Table~\ref{table:2}. These will be more or less like those of the resolved galaxies expected from ASKAP WALLABY, in terms of the spatial ($20$\--$60$\arcsec) and spectral ($4$ \kms) resolution. In this respect, the LVHIS sample galaxies have also been used for testing the performance of ASKAP WALLABY kinematics pipelines like {\sc fat} and {\sc 2dbat} (\citealt{2015MNRAS.452.3139K}; \citealt{2018MNRAS.473.3256O}).

\subsection{The data cube} \label{mp1}

We use the H{\sc i} data cubes that were reduced using `natural' weighting of the uv-data with a velocity spacing of 4 \kms. The final cubes provide angular resolution less than $\sim$40\arcsec\ ($\sim$1 kpc at $\sim$5 Mpc) and rms sensitivity of $\sim$1.5 \mjybeam\, per channel. We refer to \cite{2018MNRAS.478.1611K} for full descriptions of the calibration and imaging of the data cubes.

\begin{figure*}
\includegraphics[angle=0,width=1.0\textwidth,bb=-50 390 650 760, clip=]{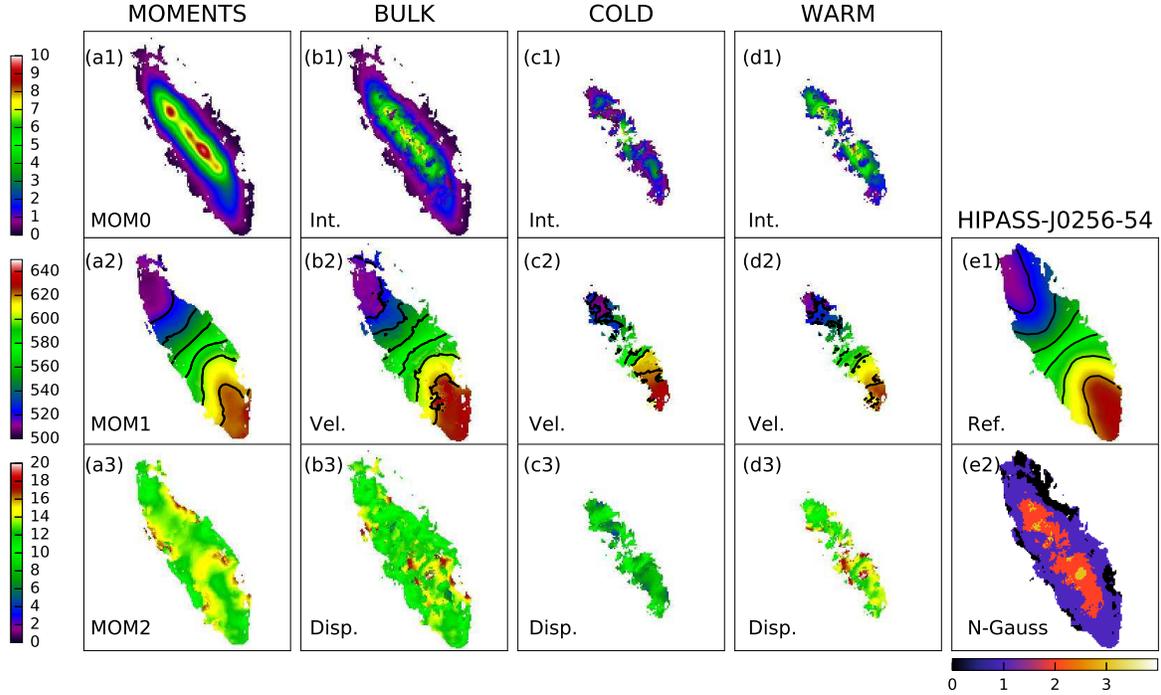}
\scriptsize
\caption{{\bf (a1\--3) \-- (d1\--3)}: 2D maps of the Gaussian components classified as bulk motions, kinematically cold or warm components of HIPASS\,J0256-54 together with its moment maps: 1) Int\-- integrated intensity in units of $10^{3}\times\rm Jy\,\rm beam^{-1}\,\rm km\,\rm s^{-1}$; 2) Vel.\-- central velocity in units of \ensuremath{\mathrm{km}\,\mathrm{s}^{-1}} and 3) Disp.\-- velocity dispersion in units of \ensuremath{\mathrm{km}\,\mathrm{s}^{-1}}. The contours overlaid on the velocity fields are spaced by 20 \kms. {\bf (e1)}: The model reference velocity field used for extracting the bulk motions. {\bf (e2)}: The optimal number of Gaussian components derived for each spaxel whose S/N values are larger than 3. See Section~\ref{mp2} for more details.
\label{Fig6}}
\end{figure*}

\begin{figure*}
\includegraphics[angle=0,width=1.0\textwidth,bb=-50 390 650 760, clip=]{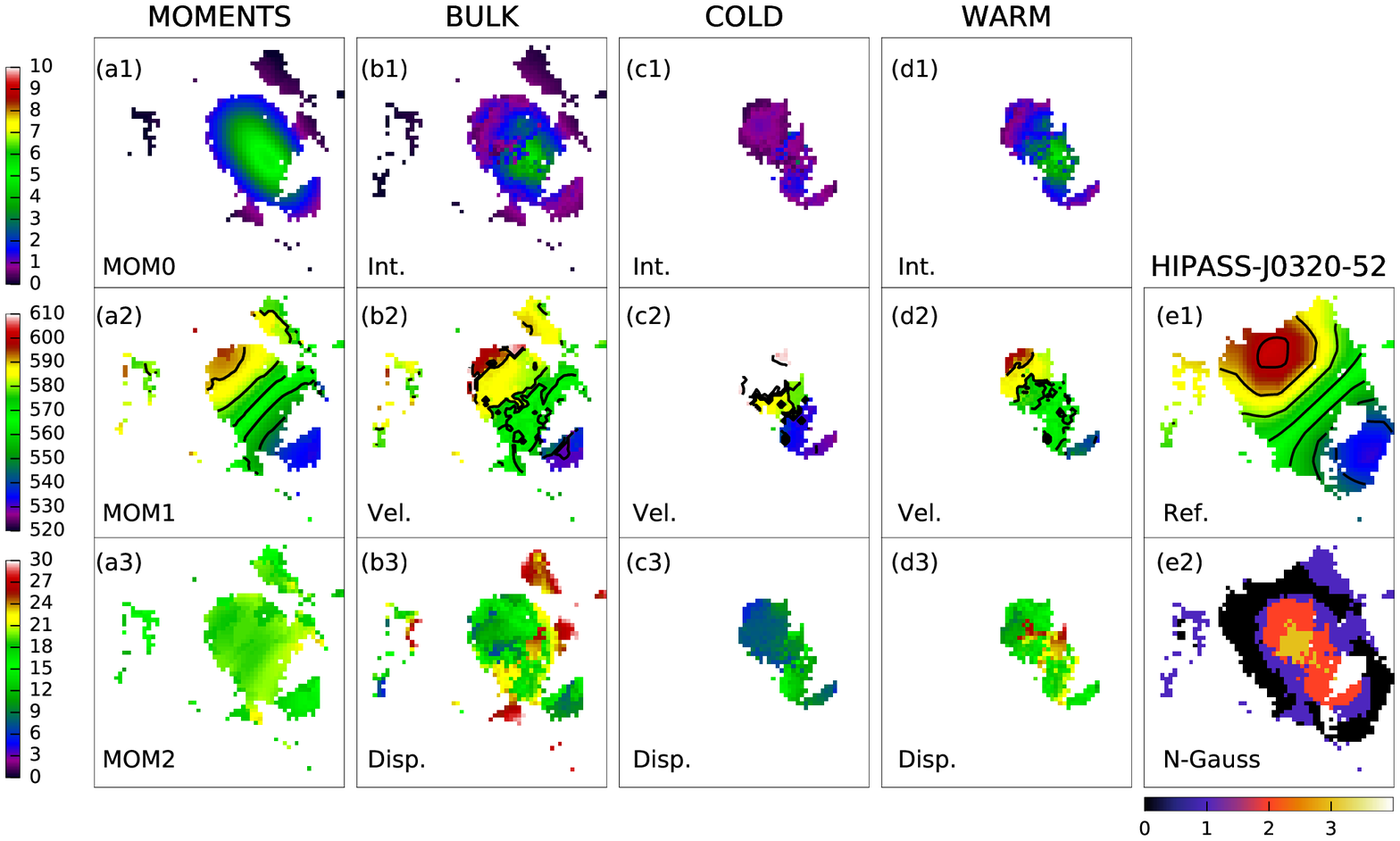}
\scriptsize
\caption{{\bf (a1\--3) \-- (d1\--3)}: 2D maps of the Gaussian components classified as bulk motions, kinematically cold or warm components of HIPASS\,J0320-52 together with its moment maps: 1) Int\-- integrated intensity in units of $10^{3}\times\rm Jy\,\rm beam^{-1}\,\rm km\,\rm s^{-1}$; 2) Vel.\-- central velocity in units of \ensuremath{\mathrm{km}\,\mathrm{s}^{-1}} and 3) Disp.\-- velocity dispersion in units of \ensuremath{\mathrm{km}\,\mathrm{s}^{-1}}. The contours overlaid on the velocity fields are spaced by 10 \kms. {\bf (e1)}: The model reference velocity field used for extracting the bulk motions. {\bf (e2)}: The optimal number of Gaussian components derived for each spaxel whose S/N values are larger than 3. See Section~\ref{mp2} for more details.
\label{Fig7}}
\end{figure*}

\begin{figure*}
\includegraphics[angle=0,width=1.0\textwidth,bb=-50 390 650 760,clip=]{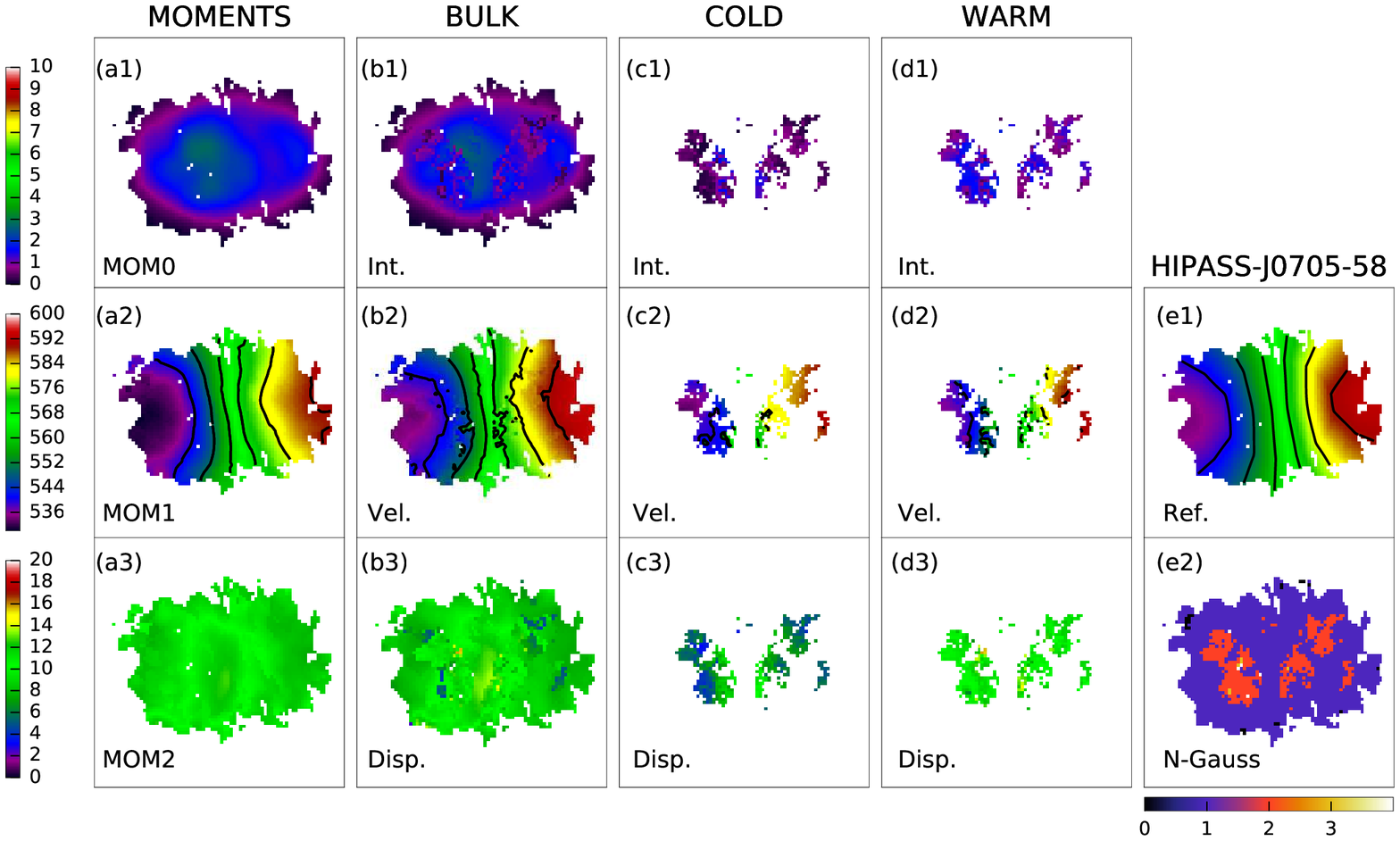}
\scriptsize
\caption{{\bf (a1\--3) \-- (d1\--3)}: 2D maps of the Gaussian components classified as bulk motions, kinematically cold or warm components of HIPASS\,J0705-58 together with its moment maps: 1) Int\-- integrated intensity in units of $10^{3}\times\rm Jy\,\rm beam^{-1}\,\rm km\,\rm s^{-1}$; 2) Vel.\-- central velocity in units of \ensuremath{\mathrm{km}\,\mathrm{s}^{-1}} and 3) Disp.\-- velocity dispersion in units of \ensuremath{\mathrm{km}\,\mathrm{s}^{-1}}. The contours overlaid on the velocity fields are spaced by 10 \kms. {\bf (e1)}: The model reference velocity field used for extracting the bulk motions. {\bf (e2)}: The optimal number of Gaussian components derived for each spaxel whose S/N values are larger than 3. See Section~\ref{mp2} for more details.
\label{Fig8}}
\end{figure*}

\begin{figure*}
\includegraphics[angle=0,width=1.0\textwidth,bb=-50 390 650 760 ,clip=]{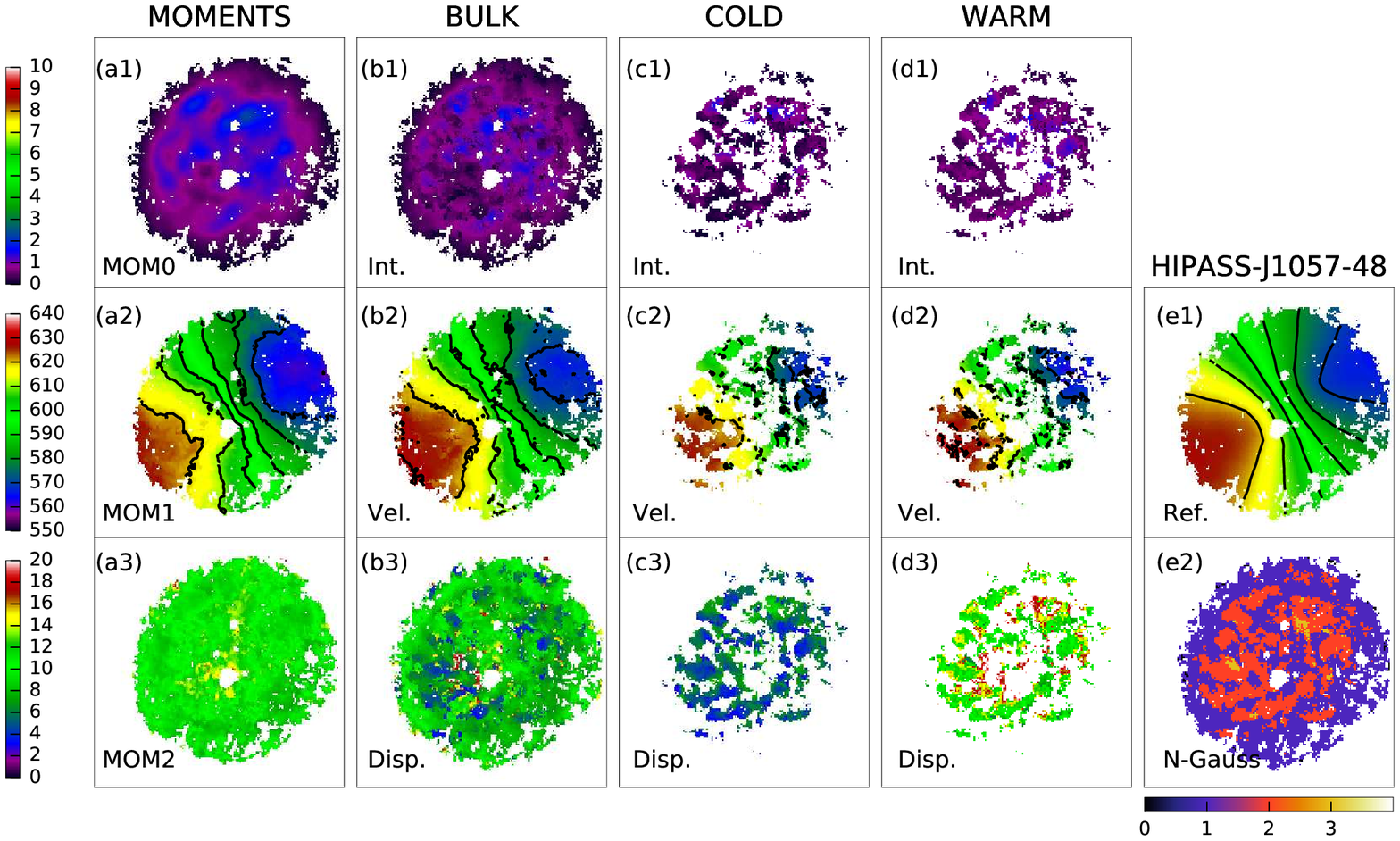}
\scriptsize
\caption{{\bf (a1\--3) \-- (d1\--3)}: 2D maps of the Gaussian components classified as bulk motions, kinematically cold or warm components of HIPASS\,J1057-48 together with its moment maps: 1) Int\-- integrated intensity in units of $10^{3}\times\rm Jy\,\rm beam^{-1}\,\rm km\,\rm s^{-1}$; 2) Vel.\-- central velocity in units of \ensuremath{\mathrm{km}\,\mathrm{s}^{-1}} and 3) Disp.\-- velocity dispersion in units of \ensuremath{\mathrm{km}\,\mathrm{s}^{-1}}. The contours overlaid on the velocity fields are spaced by 10 \kms. {\bf (e1)}: The model reference velocity field used for extracting the bulk motions. {\bf (e2)}: The optimal number of Gaussian components derived for each spaxel whose S/N values are larger than 3. See Section~\ref{mp2} for more details.
\label{Fig9}}
\end{figure*}

\begin{figure*}
\includegraphics[angle=0,width=1.0\textwidth,bb=-50 390 650 760 ,clip=]{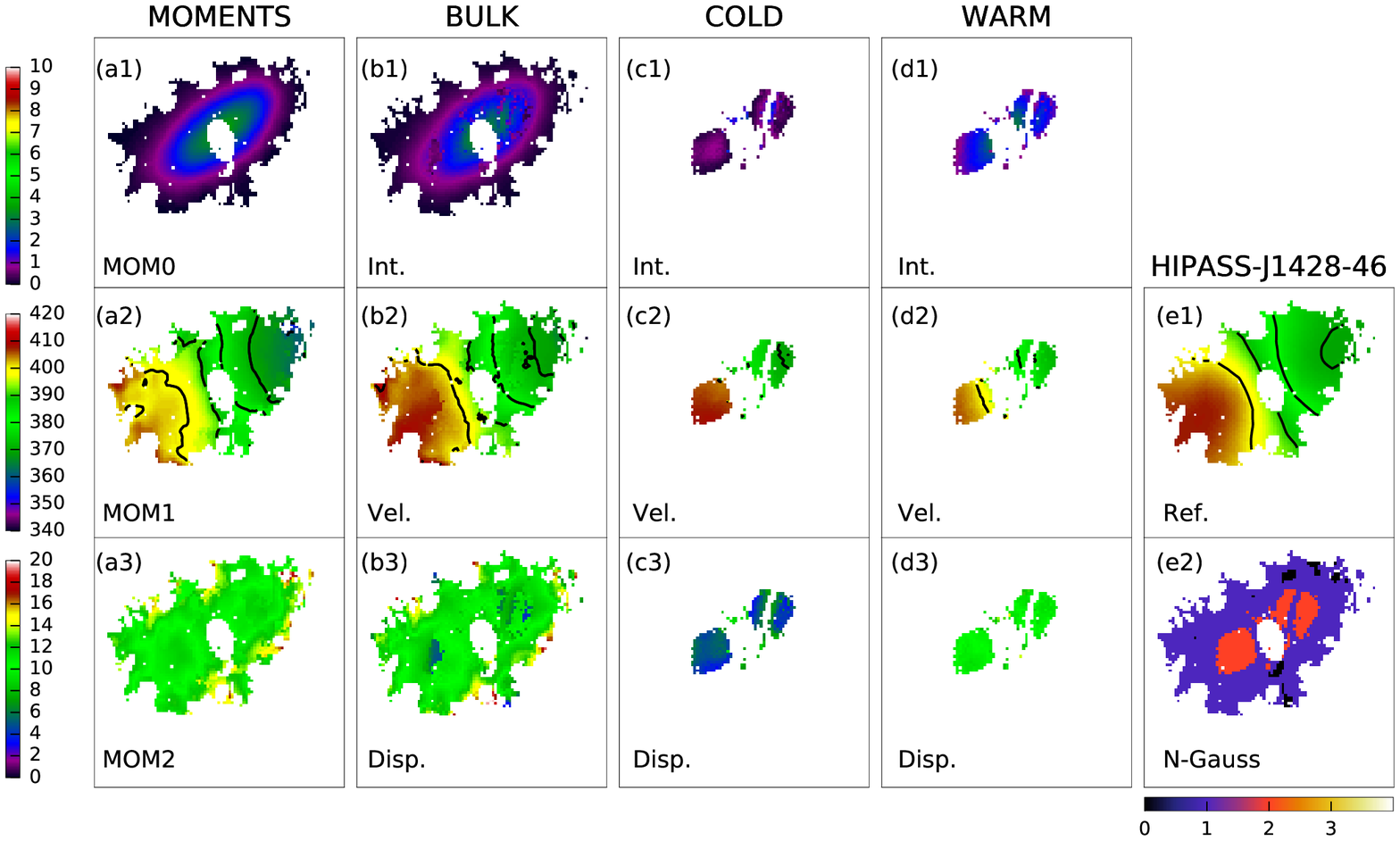}
\scriptsize
\caption{{\bf (a1\--3) \-- (d1\--3)}: 2D maps of the Gaussian components classified as bulk motions, kinematically cold or warm components of HIPASS\,J1428-46 together with its moment maps: 1) Int\-- integrated intensity in units of $10^{3}\times\rm Jy\,\rm beam^{-1}\,\rm km\,\rm s^{-1}$; 2) Vel.\-- central velocity in units of \ensuremath{\mathrm{km}\,\mathrm{s}^{-1}} and 3) Disp.\-- velocity dispersion in units of \ensuremath{\mathrm{km}\,\mathrm{s}^{-1}}. The contours overlaid on the velocity fields are spaced by 10 \kms. {\bf (e1)}: The model reference velocity field used for extracting the bulk motions. {\bf (e2)}: The optimal number of Gaussian components derived for each spaxel whose S/N values are larger than 3. See Section~\ref{mp2} for more details.
\label{Fig10}}
\end{figure*}

\begin{figure*}
\includegraphics[angle=0,width=1.0\textwidth,bb=-50 390 650 760 ,clip=]{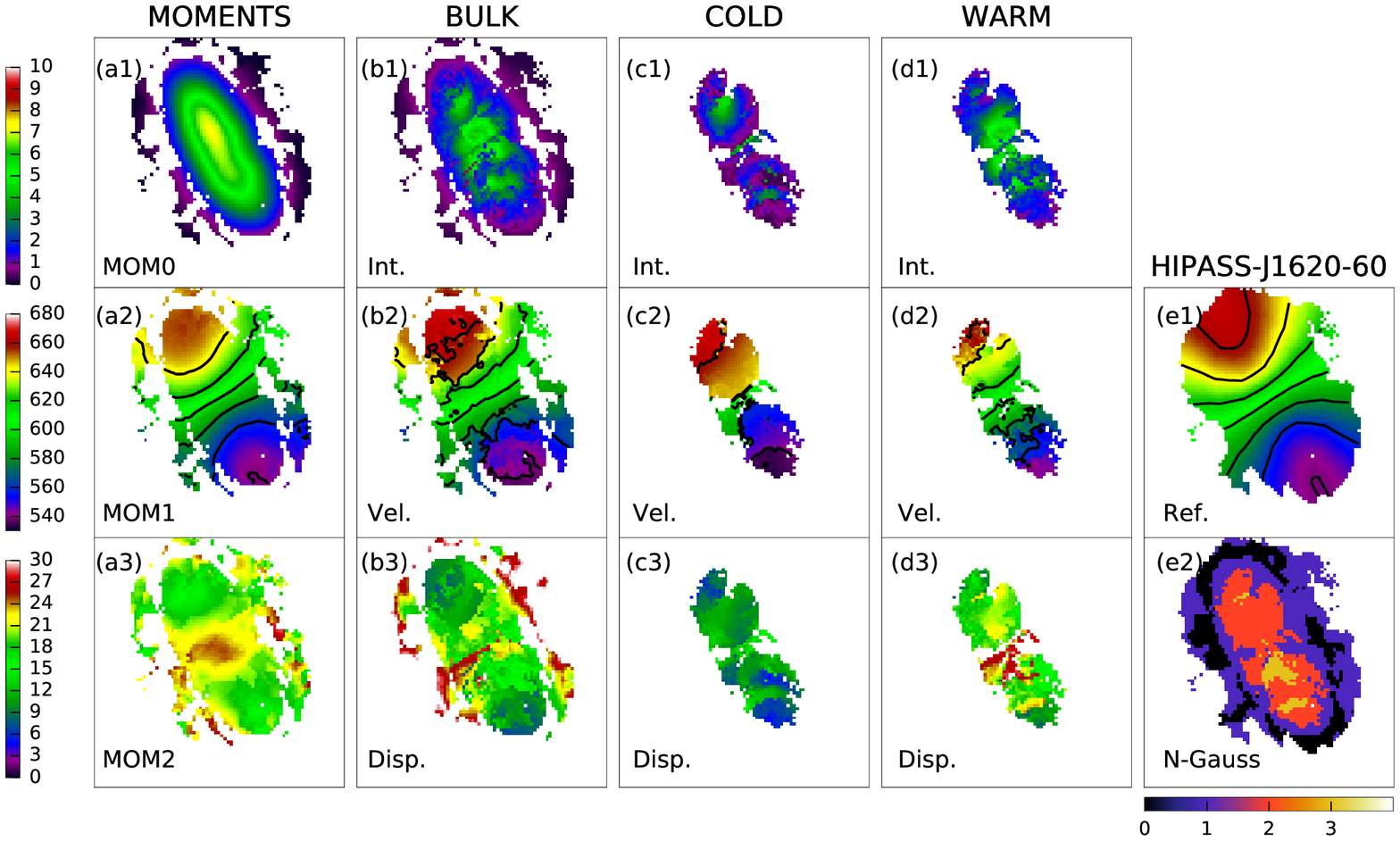}
\scriptsize
\caption{{\bf (a1\--3) \-- (d1\--3)}: 2D maps of the Gaussian components classified as bulk motions, kinematically cold or warm components of HIPASS\,J1620-60 together with its moment maps: 1) Int\-- integrated intensity in units of $10^{3}\times\rm Jy\,\rm beam^{-1}\,\rm km\,\rm s^{-1}$; 2) Vel.\-- central velocity in units of \ensuremath{\mathrm{km}\,\mathrm{s}^{-1}} and 3) Disp.\-- velocity dispersion in units of \ensuremath{\mathrm{km}\,\mathrm{s}^{-1}}. The contours overlaid on the velocity fields are spaced by 20 \kms. {\bf (e1)}: The model reference velocity field used for extracting the bulk motions. {\bf (e2)}: The optimal number of Gaussian components derived for each spaxel whose S/N values are larger than 3. See Section~\ref{mp2} for more details.
\label{Fig11}}
\end{figure*}

In the panels (a1-3) of Figs.~\ref{Fig6} to \ref{Fig11}, we show the moment maps of the galaxies extracted from the \hi data cube using the standard method with simple clipping. As discussed earlier, the conventional moment analysis is limited in quantifying non-Gaussian profiles and determining their representative properties, often giving estimates which are biased by noise or complex velocity structure. The galaxies' complex \hi velocity structure could result in the distorted velocity pattern of the moment 1 images shown in Figs.~\ref{Fig6} to \ref{Fig11}. This could partly be caused by star formation processes in the galaxies. These processes could interact with the ISM in a way that clearly disturbs the ambient ISM, resulting in distorted iso-velocity contours. In the following sections, we apply our profile decomposition technique to the data cubes of the sample galaxies, in order to improve the kinematic analyses.

\subsection{Profile decomposition} \label{mp2}

We run the code on the \hi data cubes of the 6 LVHIS sample galaxies to perform a Bayesian profile decomposition from which the best-fitting models for individual line profiles are derived. As in the case of the artificial data cubes discussed in Section~\ref{pd1}, we let the code fit a line profile with up to five Gaussian components, and find the best model for each line-of-sight profile in accordance with the computed BIC statistics. 

\subsubsection{Quantification of complex \hi structure and kinematics} \label{mp2-1}
Maps of the optimal number of Gaussian components derived for each spaxel whose S/N values are larger than 3 are shown in the panel (e2) of Figs.~\ref{Fig6} to \ref{Fig11}. There are line profiles which are best fitted by a single Gaussian with $m$=1 in Eq.~\ref{eq:1}, distributed across the galaxies. Aside from the line profiles with low S/N values, some of these single Gaussian profiles could be the leftover ISM from gas outflows driven by star formation or SNe explosions. The pixel values of the maps in the panel (e2) of Figs.~\ref{Fig6} to \ref{Fig11} range from 1 to 3, with a median value of 2 for most galaxies. This indicates that there is a significant fraction of non-Gaussian profiles in the data cubes. The extracted maps will be useful in quantifying the complexity of H{\sc i} structure as well as kinematics of the galaxies.

\subsubsection{Kinematically cold and warm components} \label{mp2-2}
The kinematically cold and warm \hi components separated with respect to their velocity dispersions are mapped in the panels (c1\--3) and (d1\--3) of Figs.~\ref{Fig6} to \ref{Fig11}. As discussed in Section~\ref{pd2}, we choose the ones with the narrowest and widest velocity dispersions as the kinematically cold and warm components, respectively, to perform the Gaussian decomposition. This classification only holds for the spaxels whose optimal number of Gaussian components found to be larger than 2.

In the comparison of the velocity fields of the cold and warm components, they do not appear to be much different except for some localised regions. In these regions, the difference of velocity dispersion between the cold and warm components is likely to be enhanced as shown in the panels (c3) and (d3) of the figures. In addition, the warm components tend to be more concentrated towards the central region of some galaxies like HIPASS\,J0256-54, HIPASS\,J0320-52 and HIPASS\,J1620-60 as shown in their integrated intensity maps (c1 and d1 panels). This could be caused by hydrodynamical processes resulting from star formation activity in the galaxies. The kinematically cold components may be acting as a gas reservoir in the formation of molecular gas which correspondingly fuels star formation. Meanwhile, the warm components may be associated with the deposition of energy from on-going star formation in the galaxies. More quantitative analysis and discussion on the pixel-by-pixel comparison of multi-wavelength data of the galaxies including our decomposed maps will be presented in the subsequent paper (Oh et al. in prep.).

\subsubsection{Bulk motions} \label{mp2-3}
From the Gaussian components for each galaxy, we extract the ones moving at the closest velocities to the reference velocity field. These are most likely to trace the global kinematics of the galaxy. We then classify the rest of the Gaussian components deviating from the bulk motions as non-circular motions.

For the reference velocity field, we construct a model velocity field using the fit results from {\sc 2dbat} in \cite{2018MNRAS.473.3256O}. As discussed in \cite{2018MNRAS.473.3256O}, {\sc 2dbat} uses {\it basis spline} functions \citep{1978pgts.book.....D}, the so-called `B-splines' to regularise any radial variations of $\phi$ and $i$.
The degree of regularisation for each galaxy is manually controlled by changing the order of the B-splines, depending on the $\rm N^{\rm semi-mx}$ and the level of radial variation of the ring parameters. We refer to \cite{2018MNRAS.473.3256O} for the full description of {\sc 2dbat} and its performance test. 

In the panel (e1) of Figs.~\ref{Fig6} to \ref{Fig11}, the model reference velocity fields derived using {\sc 2dbat} are presented. These are used for classifying the decomposed Gaussian components in the H{\sc i} data cubes of the galaxies. The bulk motion maps of the 6 LVHIS sample galaxies extracted assuming their reference velocity fields are shown in the panels (b1-3) of Figs.~\ref{Fig6} to \ref{Fig11}. 
As a comparison, in the panel (a2) of the figures, we show the intensity-weighted mean (IWM) velocity fields (moment 1) of the galaxies. 
In general, the bulk and IWM velocity fields are not much different with each other except for some localised regions where multiple kinematic components are present as quantified in the maps of the optimal number of Gaussian components in panel (e2). Despite the local wiggles in the iso-velocity contours, the kinematic components moving at velocities close to the underlying kinematics predicted by the reference velocity fields are extracted in the bulk velocity fields.

We re-iterate that all the decomposed Gaussian components in this test satisfy our decomposition criteria including that 1) the S/N value of the peak flux is larger than 3, and 2) the dispersion is larger than one channel resolution. We also emphasise that the bulk velocity field of a galaxy extracted using our method is dependent on the assumed model reference velocity field, and should only be considered as representative of the kinematics for the circularly rotating component of the galaxy. More detailed kinematics analysis using the extracted bulk motions of the galaxies will be discussed in the companion paper (Oh et al. in prep.).

\section{Summary} \label{conc}

In this paper, we present a new algorithm for the analysis of velocity profiles of data cubes from observations of \hi, or other spectral lines, based on a Bayesian MCMC technique. The algorithm deals with multiple Gaussian components to find a model which best describes the overall line profile, including non-Gaussianity and asymmetry. For each line profile, it firstly fits a series of models comprising different numbers of Gaussian components ranging from 1 up to the maximum number supplied by the user. Among the competing models, it then finds the best model in terms of their BIC values, these are calculated from the derived log-likelihood estimate for each model. This is particularly useful for the analysis of asymmetric non-Gaussian velocity profiles, characteristic of the ISM of many galaxies with complex structure and kinematics.

Compared to the standard $\chi^{2}$ minimisation technique, the Bayesian fits via MCMC sampling are less sensitive to initial priors, and robust at finding the true global best-fit parameters. Apart from specifying a range for the priors of each parameter, the analysis is fully automated. The decomposed Gaussian components can also be classified by: (1) their velocity deviation against a reference circular rotation (e.g.\ to separate bulk motions from non-circular motions); or (2) their velocity dispersion (e.g.\ to separate cold and warm components). This classification step can be further improved by adopting a new reference velocity field which is derived from a kinematic analysis made using the extracted bulk velocity from the previous step.

Code written in C and MPI which implements the above algorithm is freely available. To check reliability and robustness, we test the code using 52 artificial \hi data cubes built using the observational properties of dwarf, intermediate-mass, and massive galaxies as presented in \cite{2015MNRAS.452.3139K}. The model cubes cover a wide range of observational parameters expected from a large survey, in terms of the galaxy geometry (the size, inclination, position angle and scale height) and kinematics (rotation curves, velocity dispersion and angular momentum vector) at various S/N values. We performed the profile decomposition of the model galaxies using the code in a fully automated manner with the reference velocity fields which were derived using {\sc 2dbat}. 

For the model galaxies inclined less than $60^{\circ}$ and resolved by more than four beams across their major axes as well as satisfying the decomposition criteria adopted in our test such as the S/N value of the peak flux ($> 3$) and the velocity dispersion limit ($> 1$ channel resolution), the code was able to recover the input Gaussian components correctly. However, if the galaxies were affected by either beam smearing or projection effects, then this leads to bias in the derived Gaussian parameters. From this, we conclude that the algorithm provides reliable profile decomposition for well-resolved ($\rm N^{\rm semi-mx} > 4$) galaxies with inclinations less than $60^{\circ}$ as long as the lowest S/N value of the decomposed Gaussian components is larger than 3.   

As a practical application, we apply the technique to \hi data cubes of 6 sample galaxies taken from the ATCA LVHIS survey. The optimally decomposed Gaussian components of the galaxies are then classified kinematically as cold, warm or bulk components with respect to their reference velocity fields derived using {\sc 2dbat}.  
We find that the cold and warm components of the galaxies differ little in their velocity fields, but the warm components tend to be pronounced in some localised regions of the integrated intensity and velocity dispersion maps. This could be caused by hydrodynamical processes in these regions like stellar winds, SNe explosions etc.

To derive systematic bulk motions within the galaxies, we construct their reference velocity fields using the fit results from 2D tilted-ring analysis made with {\sc 2dbat}. We then extract the Gaussian components which are closest to the velocities predicted by the reference velocity fields. The bulk velocity fields are, in general, consistent with the moment 1 maps except for the regions where multiple kinematic components are present. However, the underlying kinematics of the galaxies are expected to be better traced by their bulk velocity fields, which separate any random motions from the circular rotation.

According to our tests using both artificial and real galaxies, the profile decomposition algorithm will be useful for quantifying the complexity of ISM structure and kinematics, and investigating its relationship with hydrodynamical processes and mass distribution in galaxies detected in future large-scale spectral-line surveys.

\section*{Acknowledgements}

The profile decomposition with the parallel implementation of the code was performed by using a high performance computing cluster, `POLARIS' at Korea Astronomy and Space Science Institute (KASI). Parts of this research were conducted by the Australian Research Council Centre of Excellence for All Sky Astrophysics in 3 Dimensions (ASTRO 3D) through project number CE170100013.

\bibliography{ms}

\end{document}